# Scaling Laws for NanoFET Sensors

Fu-Shan Zhou, and Qi-Huo Wei

*Liquid Crystal Institute, Kent State University, Kent, OH 44242*
Email: qihuo@lci.kent.edu; Dated : Oct. 5, 2007

The sensitive conductance change of semiconductor nanowires and carbon nanotubes in response to binding of charged molecules provide a novel sensing modality which is generally denoted as nanoFET sensors. In this paper, we study the scaling laws of nanoplate FET sensors by simplifying nanoplates as random resistor networks with molecular receptors sitting on lattice sites. Nanowire/tube FETs are included as the limiting cases where the device width goes small. Computer simulations show that the field effect strength exerted by the binding molecules has significant impact on the scaling behaviors. When the field effect strength is small, nanoFETs have little size and shape dependence. In contrast, when the field-effect strength becomes stronger, there exists a lower detection threshold for charge accumulation FETs and an upper detection threshold for charge depletion FET sensors. At these thresholds, the nanoFET devices undergo a transition between low and large sensitivities. These thresholds may set the detection limits of nanoFET sensors, while could be eliminated by designing devices with very short source-drain distance and large width.

The recent development of semiconductor nanomaterials including nanowires, nanobelts and carbon nanotubes has attracted considerable attention because of their fascinating potential applications in a new generation of nanoelectronics, nanophotonics and nanosensors [1-7]. This is particularly the case for nanosensors made of semiconductor nanowires (NWs) and carbon nanotubes (NTs). When these semiconductor NWs or NTs are chemically or biologically functionalized with molecular receptors, specific binding of charged analyte molecules results in depletion or accumulation of charge carriers and a change in conductance. This effect is generally denoted as nanoFET or bioFETs [8-12]. These chemical/biological field effects have been employed for designing NW/NT sensors for various chemical and biological detections [10-18].

These nanoFET biosensors are commonly believed to be ultra-sensitive with potential single molecule sensitivity considering that the depletion and accumulation of charge carriers can affect the entire cross-sectional conduction pathway of these devices. The experimental results from different groups, however, show diverse ranges of detection limits and sensitivities [10, 14, 16, 17, 19, 20]. This is understandable considering that the sensitivity, detection limit, and dynamic range of such nanoFET sensors should depend on device parameters such as charge carrier density, receptor density, device geometry and dimension. These size dependent effects, or the device scaling laws, are critical to practical applications. Since the concentrations of analyte molecules could vary significantly, it is desirable to be able to tailor the device sensitivity and dynamic range in accordance with the concentrations of analyte molecules. The ultra-low detection limit of these nanoFETs is critically required for the detection of trace amount of analyte molecules for applications in detections of chemical and biological threats such as explosives and protein biotoxins. However, so far little has been done or understood in these aspects of the scaling laws.

Recent analyses on diffusion-limited binding processes for analyte molecules have shown that NWs are superior to planar FET sensors in low analyte concentrations because they take less time to accumulate detectable amount of molecules [21, 22]. The diffusion-limited binding process, however, does not set the absolute limit for nanoFET sensors, since it can be circumvented by continuously flowing the analyte solutions or employing other microfluidic mixing techniques [23] or sample preconcentration techniques [24].

In this paper, we address how the sensitivity and detection limit of nanoplate FET sensors depend on their geometrical parameters. By simplifying the nanoplates as resistor networks, the scaling law problem is converted into finite size effects of a random resistor network. The ratio ($\Delta$) of conductance (or resistance) increase of resistor bonds neighboring to the molecular binding site is utilized to characterize the field effect strength. Simulation results show that the field effect strength $\Delta$ exerted by the binding molecule has substantial effects on the scaling behaviors. When the field effect strength is small ($\Delta<10$), nanoFET devices show no size and shape dependence. While when the field-effect strength is increased ($\Delta>10$), there exist a lower detection threshold for charge accumulation nanoFETs and an upper detection threshold for charge depletion nanoFETs. At the lower detection threshold, the nanoFET devices undertake transitions from low to dramatically high sensitivities. At the upper threshold, the nanoFET devices undertake transitions from high to very low sensitivities. These thresholds could set the detection limit for the nanoFET devices. To surmount these detection limits, a new configuration of nanoFET is proposed with very short source-drain distance and large width.

We start with a semiconductor nanoplate of small thickness (down to ~10nm) whose surfaces are covered with a thin oxide layer, and assume that receptor molecules are periodically immobilized on the oxide surface (Fig. 1a). This functionalized nanoplate can be approximated by a square lattice network of resistors with the receptors sitting on lattice sites (Fig. 1b). As in experiments, the electron transport in the devices is assumed to be diffusive and the devices are working in the linear regime (i.e., low source-drain voltage). In another word, the conductance of the resistor bond does not vary with the voltage applied on it. Also it is important to point out that the gate voltage other than those imposed by binding molecules is not taken into account for simplicity. Random resistor networks have been successfully employed to explain a number of phenomena in continuum media, especially the electrical transport properties in systems of compact mixtures of conducting and non conducting materials or homogeneous two-phase systems with one phase more conductive than the other [25, 26]. Applications of the random resistor networks here should also capture the essential features of the scaling laws of nanoFET sensors.

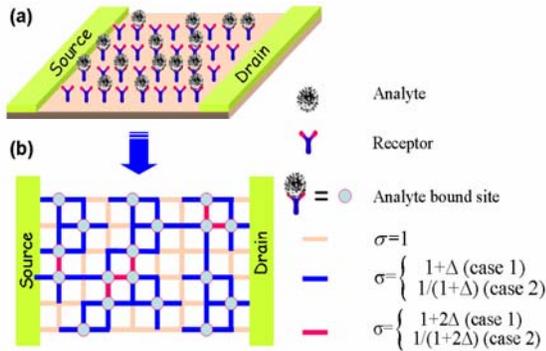

**Figure 1.** Schematic of the resistor network model: a nanoplate FET sensor (a) is simplified as a resistor network with receptor molecules sitting at lattice sites (b). Orange bonds: no binding at both lattice sites; blue bonds: one site is bound by an analyte molecule; red bonds: both sites bound by analyte molecules.

Although it is not well known how the receptors are immobilized on real device surface, the scaling behaviors observed in our model should apply generally. NW/NT FETs can be considered as the limiting cases where the width of the resistor networks becomes very small ($W$ ~ a few lattice constants). It is worth pointing out that the receptors and analytes, though represented as antibodies and antigens in Fig.1, can be a wide range of molecules such as 3-aminopropyltriethoxysilane (APTES) and $H^+$ ions [8], complementary DNA molecules [10, 14], tyrosine kinase and small inhibitor molecules [19], and so on.

Two different cases need to be specifically distinguished. The first case is the conductance increase upon the binding of analyte molecules, which encompasses positively charged analyte molecules on n-type nanoFETs and negatively charged analyte molecules on p-type nanoFETs. The binding of analyte molecules carrying charges opposite to the main carriers in the FET leads to accumulation of charge carriers in regions under the binding molecules. The second case refers to a resistance increase upon molecular binding, which corresponds to positively charged analyte molecules on p-type nanoFETs or negatively charged analyted molecules on n-type nanoFETs. The same sign of analyte molecular charges and the main carriers in the FET leads to depletion of main carriers beneath the binding molecules. In our simulations, we do not distinguish p- and n-type nanoFETs, rather we consider only conductance or resistance increases in general.

Without adsorption of analyte molecules, bonds of the resistor network have a uniform conductance that is assumed to be 1 unit. When a charged analyte molecule is specifically bound to one receptor molecule, an electric field is exerted both inside and outside the nanoplate. The electric field outside the plate is primarily screened by counter ions in the solution, and therefore the region inside the plate affected by this field is approximately of the size of one Debye screening length. Full details of these field effects necessitate numerical solutions of the non-linear Poisson-Boltzmann (PB) equations both inside [27] and outside the semiconductor plate [28]. Here, we try to catch the essential physical picture of the scaling laws without going to detailed solutions for the PB equations, and reasonably assume that the field effects are represented by the conductance change of four nearest bonds of the bound receptors in the resistor networks.

In order to characterize the field effect, we define a dimensionless parameter $\Delta$ which is the ratio of bond conductance increase for the first case (i.e., charge accumulation) and the ratio of bond resistance increase for the second case (charge depletion). When two neighboring sites are occupied by analyte molecules, the local electric field between them is roughly the superposition of two charged molecules, and therefore the conductance change of the bond connecting these two sites is assumed to be doubled ($2\Delta$) (Fig. 1).

The probability $p$ that a receptor (denoted as R) is bound by an analyte molecule (denoted as A) is related to the analyte concentration. The binding reaction R+A→R$^*$, as a first order approximation, can be described by the Langmuir kinetic law [29]:

$$dp/dt = k_+ c_A (1-p) - k_- p, \qquad (1)$$

Where R$^*$ denotes the receptor molecule bound by an analyte molecule, $p$ is also the surface concentration of bounded receptors ($R^*$), $c_A$ is the analyte concentration in the immediate vicinity of the surface, $1-p$ is the surface concentration of unbounded receptors ($R$), and $k_+$ and $k_-$ are the rate coefficients. When the binding reaction reaches its equilibrium: $p = kc_A/(1+kc_A)$, where $k=k_+/k_-$ is the equilibrium constant. At small analyte concentrations ( $kc_A \ll 1$ ), the molecular binding probability *p* is a linear function of the analyte molecule concentration.

Numerical simulations are carried out for networks of different length (*L*) and width (*W*) ranging from 2 to $2^{10}$ lattice constants. In order to calculate the conductance of these resistor networks, a voltage is applied between the source and drain electrodes, and voltages at each lattice site are related to the voltages of their nearest sites through a set of Kirchoff equations. Direct matrix inversions are employed to solve these linear Kirchoff equations, and then the total current through the network and the total network conductance are calculated accordingly. In addition, no periodical boundary conditions are used in our simulations. Representative simulation results of the conductance in the first and second cases are presented in Fig. 2 and 3 respectively as a function of the binding probability for different device widths and lengths and for three representative values of *Δ*. The conductance *Σ* has been normalized by the original conductance of the network before molecule binding, and is dimensionless.

For small *Δ*(<10), little length and width dependence is found for both cases (Fig. 2a-b; Fig. 3a-b). In other words, the normalized network conductance is approximately a universal function of the binding probability for all device sizes and shapes. Considering the fact that the reaction constant *k* depends on specific analyte/receptor molecules involved, this universal behavior of the resister network suggests that the experimentally measured relationship *Σ~p*, or the calibration curve, is independent of device size and shapes as long as these devices are functionalized with the same receptors at the same density and working at the same experimental conditions, especially at the same ion concentration.

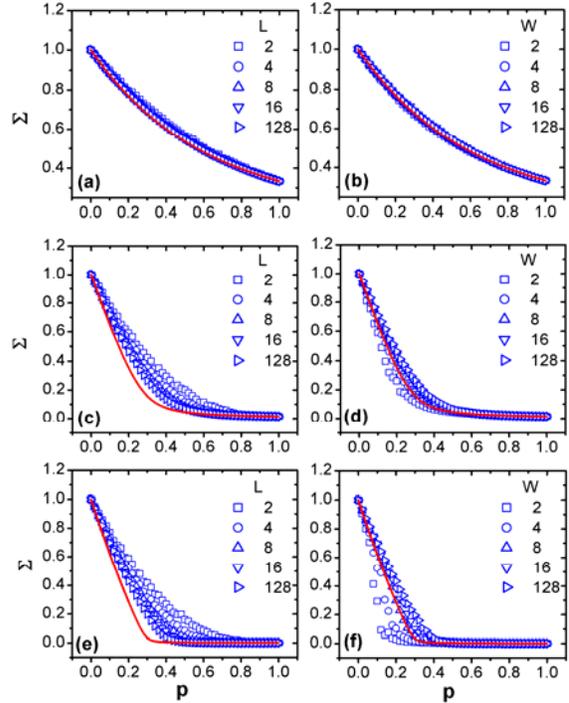

**Figure 3.** Normalized conductance of the random resistor networks as a function of binding probability *p* in resistance increase case with *Δ* equal to 2 (a, b), 64 (c, d) and 1024 (e, f). Left column: fixed width *W=256*; right column: fixed length *L=256*. For clarity, only simulation data for *L* (left) and *W* (right) varying between 2 and 128 are shown.

When the field effects become stronger (*Δ>10*) (Fig. 2c-f, Fig. 3c-f), the response of nanoFETs to molecular binding becomes more size and shape dependent. A threshold probability $p_c$ can be observed for most device parameters. For the first case, the device sensitivity defined as *dΣ/dp* increases sharply when the binding probability (or the analyte concentration) is raised above the threshold. The threshold concentration sets the lower detection limit of the device if the conductance change *dΣ/dp* below the threshold is not detectable. This lower limit of detection increases with the source-drain length *L* while decreases with the device width *W*. In contrast, the device sensitivity *dΣ/dp* for the second case decreases dramatically when the binding probability goes beyond the threshold. The threshold could set the upper detection limit of the device if

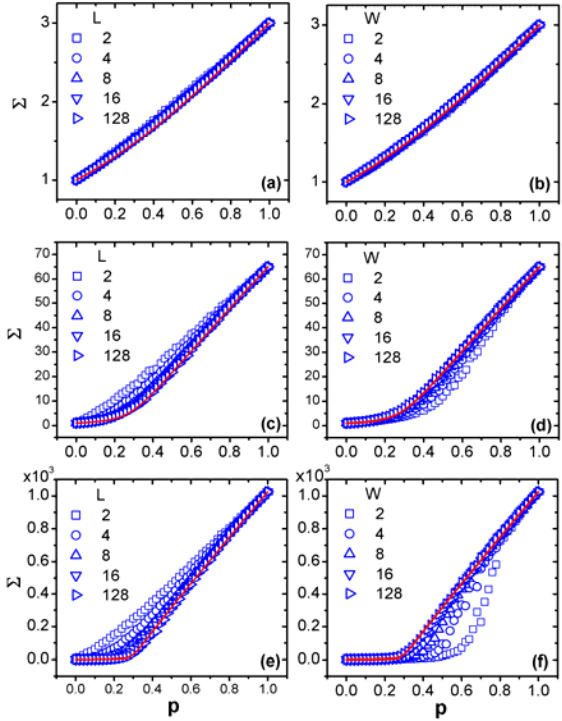

**Figure 2** Normalized conductance of the random resistor networks as a function of binding probability *p* in the conductance increase case with *Δ* equal to *2* (a, b), *64* (c, d) and *1024* (e, f). Left column: fixed width *W*=256; right column: fixed length *L*=256. For clarity, only simulation data for L (left) and W (right) varying between 2 and 128 are shown.

$d\Sigma/dp$ for $p$ above the threshold probability is not detectable. In contrast to the first case, this upper threshold decreases with the source-drain length $L$ while increases with the device width $W$. As a result, sensors of the second case are more suitable for detection of extremely low concentration of analytes, while sensors of the first case are more suitable for relatively high concentration of analytes. The physics behind these seemingly counterintuitive scaling behaviors can be easily understood by looking at two extreme situations. One is close to the NW/NT FET sensors where the device width $W=1$ (Fig. 4a). The device can be considered as composed of conductors $\sigma_i$ in serial connection, and the total conductance $\Sigma$ can be expressed as: $1/\Sigma \sim \sum_i 1/\sigma_i$. It can be expected that for large $\Delta$ in the first case (charge accumulation), $\Sigma$ is insensitive to the conductance increase $\sigma_i \rightarrow 1+\Delta$ induced by molecular binding as long as the binding probability $p$ is smaller than 1 (see blue triangles in Fig. 4c). In another word, the lower threshold $p_c$ is 1 or close to 1. While in the charge depletion case, $\Sigma$ is extraordinarily sensitive to the conductance decrease $\sigma_i \rightarrow 1/(1+\Delta)$ induced by single molecule binding, but becomes insensitive to further binding soon after p is larger than 0 since the value of $\Sigma$ already approaches zero (see blue triangles in Fig. 4d). In another word, the upper threshold $p_c$ here is 0. The other extreme situation is for minute source-drain length $L=2$ (Fig. 4b). Now the device can be considered as composed of conductors $\sigma_i$ in parallel connection, and the total conductance: $\Sigma \sim \sum_i \sigma_i$. It can be found out that for large $\Delta$ in both the charge accumulation and depletion cases, $\Sigma$ is always sensitive to the conductance increase $\sigma_i \rightarrow 1+\Delta$ or decrease $\sigma_i \rightarrow 1/(1+\Delta)$ no matter what the probability $p$ is. Computer simulations also confirmed this argument (Fig. 4c-d, red circles). This simple physics picture demonstrates the following: (1) at large field effect strength, NW/NT FET sensors indeed exhibit highest sensitivities, but that is restricted to a certain concentration range as confirmed by the simulation data in Fig. 4c-d. In another word, NW/NT FET sensors may have either an upper or a lower detection limit. (2) The new nanoFET design with very small $L$ and large $W$ should not exhibit any detection limits although its sensitivity is lower.

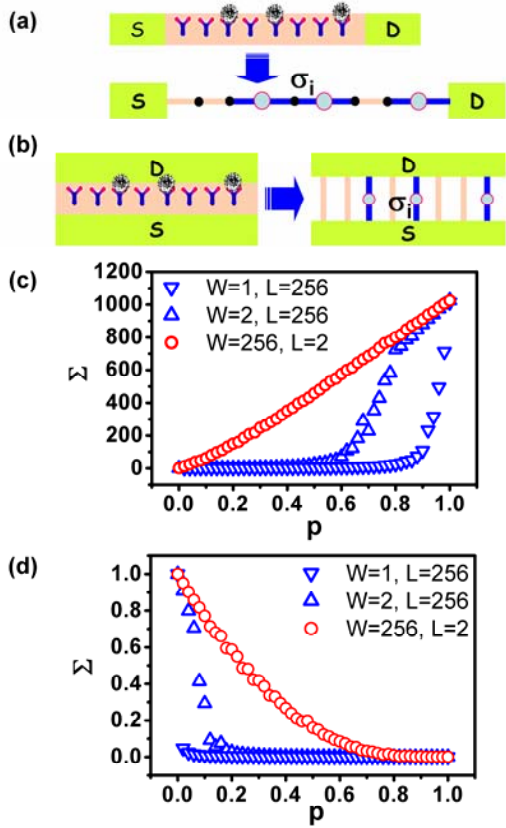

**Figure 4** Schematic of nanoFET sensors at two extreme conditions and their equivalent resistor networks: (a) the device width $W=1$ corresponding to nanowire or nanotube FET sensors; (b) new design of FET sensors with minute source-drain distance ($L=2$); (c) the simulated conductance $\Sigma$ vs. the binding probability p for the charge accumulation case for nanowires ($W=1$ and $2$, $L=256$) and the new design ($W=256$, $L=2$); and (d) the simulated conductance $\Sigma$ vs. the binding probability p for the charge depletion case for nanowires ($W=1$ and $2$, $L=256$) and the new design ($W=256$, $L=2$).

The physical origin of the thresholds is related to percolation [25, 26]. In a traditional bond percolation model, the network is composed of insulator and resistor bonds [25, 26]. When the resistor bond probability $p$ is equal to or larger than the percolation threshold $p_c$, the probability of finding a cluster of connected resistor bonds which spans over between two electrodes become non-zero, or the network transfers from an insulator to a conductor. This insulator-conductor transition is a continuous phase transition, and the percolation threshold is system size dependent. The resistor network proposed here for nanoFETs is different from the traditional one for percolation models [25, 26] in terms that the bond conductance is always nonzero (semiconductor). The resistor network in the first case can be considered as composed of "insulators" (less conductive bonds), and molecular binding causes these "insulators" to become "conductors" (more conductive bonds). While the resistor network in the second case can be considered as consisting of "conductors" (more conductive bonds), and molecular binding causes these "conductors" to become "insulators" (less conductive bonds). Therefore, for a large field effect, the thresholds observed above are corresponding to the percolation threshold $p_c$ of forming or losing a network of large conductance bonds spanning between electrodes, and the network resistance should exhibit finite size scaling similar to that in percolation transitions.

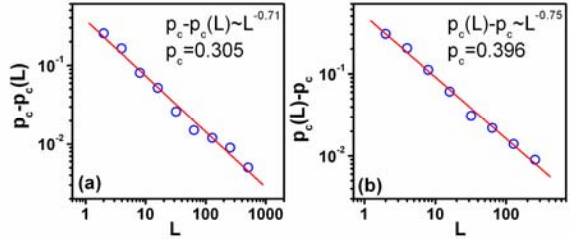

**Figure 5.** Detection limit (i.e., percolation threshold) as a function of source-drain distance $L$ for width $W=256$ for the conductance increase case (a) and the resistance increase case (b).

Finite size effects cause the shift of percolation thresholds (Fig. 3). Particularly, the percolation threshold of a system of size $L$, $p_c(L)$, behaves according to: $|p_c(L) - p_c| \sim L^{-1/\nu}$, where $p_c$ is the percolation threshold for systems of infinite size. $p_c$ in our system is found to be about 0.3 and 0.4 for the first and second cases respectively. The exponent $\nu$ is found to be $1.4 \pm 0.1$ and $1.3 \pm 0.1$ for the first and second cases respectively (Fig. 5).

Finite size scaling for percolation indicates that the variation of any property $\phi$ for a system of size $L$ obeys $\phi \sim L^{-\mu/\nu} f[L^{1/\nu}(p-p_c)]$, where the scaling function $f(x) \to 1$ when $x \to 0$; and $f(x) \to x^\mu$ when $x \to \infty$. Therefore, for a system of infinite size, $\phi \sim (p-p_c)^\mu$. Figure 6 shows the scaling plots for two representative field effect strengths $\Delta$ for both cases. For large field strengths, it can be observed that scaling plots of $\Sigma L^{\mu/\nu} \sim (p-p_c)L^{1/\nu}$ with proper exponents could lead to collapse of all data to master curves. The scaling exponent $\nu$ is found to be close to that obtained above from the detection limit plot (Fig. 5), and the scaling exponent $\mu$ is found to be around 0.91 and 1.2 for the first and the second cases (Fig. 6c-d). On the other hand, for intermediate $\Delta$, we found that this scaling law poorly applies (Fig. 6a-b).

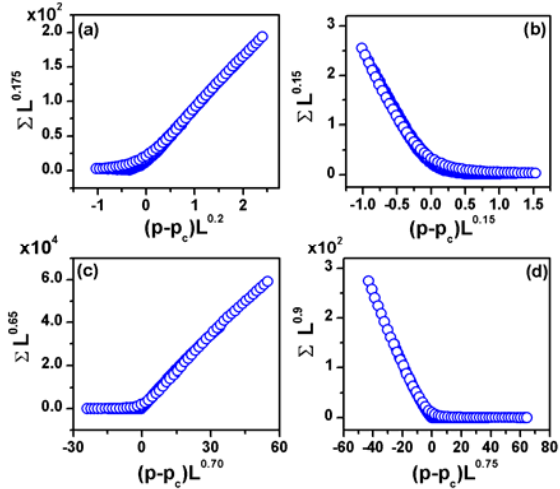

**Figure 6.** Scaling plots for two representative field effect strengths for the first (left column) and second (right column) cases with $W=256$: (a) and (b) $\Delta=64$; (c) and (d) $\Delta=1024$.

It has been known that effective media approximation (EMA) is particularly precise for the bond percolation model [25, 26, 30]. In the EMA approach, the potential drop on each bond is considered as the average potential drop obtained by replacing the random network with resistors of equal conductance ($\sigma_m$), plus a fluctuating local field whose average over large system size should go to zero. When the conductance values are distributed according to a distribution function $g(\sigma)$, the EMA leads to a condition:

$$\int d\sigma g(\sigma)(\sigma_m - \sigma)/[\sigma + (z/2 - 1)\sigma_m] = 0, \quad (2)$$

where z, the number of bonds to each lattice site, is 4 for square lattice. Our resistor network has a cubic polynomial distribution, which can be expressed for the first case as:

$$g(\sigma) = (1-p)^2 \delta(\sigma - 1) + 2p(1-p)\delta(\sigma - 1 - \Delta)$$
$$+ p^2 \delta(\sigma - 1 - 2\Delta),$$
(3)

$g(\sigma)$ for the second case can be expressed in a similar way. Combining Eq.3 and Eq.2 leads to a cubic polynomial equation for $\sigma_m$ which has only one real root solution. For large system size, EMA results fit particularly well the simulations for the first case (red solid lines in Fig. 2), while show some discrepancy for the second case at large $\Delta$ values (red solid lines in Fig. 3). The physics behind this discrepancy is not clear yet. One possible explanation is related to the fact that the bond conductance values are correlated in our model although the binding sites are random and the second case needs better corrections for this effect.

It is difficult at this moment to make a detailed comparison of our simulation results with experiments due to very limited amount of experimental work addressing the scaling laws for nanoFET sensors [17, 31]. However, our results for the first case agree qualitatively well with the recent experimental and simulation work done by Elfstrom et al [31]. Numerical results in the literature [31] clearly indicate that there exists a lower threshold of surface charge density above which the device sensitivity increases sharply (Fig. 5 in literature) and that the threshold decreases with the device width. The experimental work done by Reed and co-workers with top-down approach [17] studied the effect of nanoFET device surface areas and showed that sensitivity increases with the decrease of device surface area (mostly the device width), which is in agreement with our simulations. The devices used are about 40nm in thickness and between 50-150nm in width with source-drain length not specified, and the detection limit achieved is around 70aM [17]. The experimental work done by Lieber and co-workers using bottom-up methods has achieved 2.5fM detection limit, where their devices are mostly about 20nm in diameter and about a few μm in length [8, 14, 32]. While the work done by Z. Li and co-workers using top-down method has shown a detection limit around 20fM for DNA molecules, and their devices are about 20μm in length, 50nm in width and 60nm in thickness [10]. Although the variation of detection limits with geometrical sizes in these experiments do not infringe with our prediction, it is not certain that these experiments and our theory are in agreement since these experiments are done at different experiment conditions such as ion concentration, molecular species involved, surface receptor density and etc.

These scaling behaviors have significant impact on the performance of nanoFET sensors. The high demand for detection of trace amount chemical or biological materials necessitates designing nanoFET sensors without detection thresholds and with high sensitivity. Based on our simulations and the simple physical arguments (Fig.4), nanoFET sensors with small width and large length exhibit very high sensitivities only at certain concentration range, and may have lower detection limits in the charge accumulation case or upper detection limit in the charge depletion case. The new design proposed here to overcome these detection limits is the nanoFET sensor with very short source-drain length and large width (Fig. 4b-d). The advantages of this new design are: (1) there is no detection threshold for both charge accumulation and depletion cases; and (2) the device sensitivity is homogeneous over whole range of binding probability.

In summary, the scaling laws of nanoFET sensors are a complex problem owing to the large number of physics parameters involved, and this work is focused on the effects of geometrical parameters. Our random resistor network model and numerical studies reveal that the scaling laws vary with the field effect strength exerted by the binding molecules. For small field effect, little size and shape dependences are observed. For large field effect, there exist detection thresholds which signify transitions between sharply different sensitivities. These detection thresholds may set the detection limits of these devices. A new design with small source-drain distance and large width is proposed to surmount these detection thresholds. The problem of scaling laws for nanoFET sensors is an important one, and experimental studies on these aspects are called for.

ACKNOWLEDGMENT We thank Drs. Chia-fu Chou, Erica Forzani, Wu Lu, Jonathan Selinger, Robin Selinger, Nong-Jian Tao, Deng-Ke Yang, and Wei-Guo Yin for valuable discussions. The work is supported by a start-up fund from Kent State University.

**References:**

[1] Y. H and Lieber C M 2004 *Pure Appl. Chem.* **76** 2051
[2] Wang Z L 2004 *Mat. Today* **7** 26
[3] Yang P D 2005 *MRS Bull.* **30** 85
[4] Samuelson L 2003 *Mat. Today* **6** 22
[5] Dresselhaus M (eds) 2004 *MRS Carbon Nanotube Special Issue* 29
[6] Li Y, Qian F, Xiang J and Lieber C M 2006 *Mat. Today* **9** 18
[7] Patolsky F, Timko B P, Zheng G F and Lieber C M 2007 *MRS Bulletin* **32** 142
[8] Cui Y, Wei Q, Park H and Lieber C M 2001 *Science* **293** 1289
[9] Kong J and et al. 2000 *Science* **287** 622
[10] Li Z, Chen Y, Li X, Kamins T I, Kauka K and Williams R S 2004 *Nano Lett.* **4** 245
[11] Besteman K, Lee J O, Wiertz F G M, Heering H A and Dekker C 2003 *Nano Lett.* **3** 727
[12] Chen R J, Zhang Y G, Wang D W and Dai H J 2001 *J. Am. Chem. Soc.* **123** 3838
[13] Comini E, Faglia G, Sberveglieri G, Pan Z W and Wang Z L 2002 *Appl. Phys. Lett.* **81** 1869
[14] Hahm J and Lieber C M 2004 *Nano Lett.* **4** 51
[15] Li C and et al. 2005 *J. Am. Chem. Soc.* **127** 12484
[16] Patolsky F, Zheng G F, Hayden O, Lakadamyali M, Zhuang X W and Lieber C M 2004 *PNAS USA* **101** 14017
[17] Stern E and et al 2007 *Nature* **445** 519
[18] Kong J, Chapline M G and Dai H J 2001 *Adv. Mat.* **13** 1384
[19] Wang W, Chen C, Lin K H, Fang Y and Lieber C M 2005 *PNAS USA* **102** 3208
[20] Zheng G F, Patolsky F, Cui Y, Wang W and Lieber C M 2005 *Nat. Biotech.* **23** 1294
[21] Sheehan P E and Whitman L J 2005 *Nano Lett.* **5** 803
[22] Nair P R and Alam M A 2006 *Appl. Phys. Lett.* **88** 233120
[23] Liu R H, Lenigk R, Druyor-Sanchez R L, Yang J N and Grodzinski P 2003 *Anal. Chem.* **75** 1911
[24] Chou C F and et al. 2002 *Biophys. J.* **83** 2170
[25] Stauffer D and Aharony A 1992 *Introduction to Percolation Theory* (London: Taylor and Francis)
[26] Kirkpatrick S 1973 *Rev. Mod. Phys.* **45** 574
[27] Kankare J and Vinokurov I A 1999 *Langmuir* **15** 5591
[28] Sze S M 1985 *Physics of Semiconductor Devices* (Central Book Company)
[29] Russel W B, Saville D A and Schowalter W R 1990 *Colloidal Dispersions* (Cambridge University Press)
[30] Landauer R 1978 AIP Conf. Proc. **40** 2
[31] Elfstrom N, Juhasz R, Sychugov I, Engfeldt T, Karlstrom A E and Linnros J 2007 *Nano Lett.* (in press).
[32] Lieber C M 2007, *Trends in Anal. Chem.* **26**, IX.